# Coherent control in room-temperature quantum dot semiconductor optical amplifiers using shaped pulses

O. KARNI,[1,2]* A. K. MISHRA,[1,2] G. EISENSTEIN,[1,2] V. IVANOV,[3] J. P. REITHMAIER,[3]

[1] Technion – Isreal Institute of Technology, Electrical Engineering Dept., Haifa 32000, Israel.
[2] Technion – Isreal Institute of Technology, Russell Berrie Nanotechnology Institute, Haifa 32000, Israel.
[3] University of Kassel, Institute of Nanostructure Technologies and Analytics, Technische Physik, CINSaT, 34132 Kassel, Germany.
*Corresponding author: oulrik@tx.technion.ac.il

**We demonstrate the ability to control quantum coherent Rabi-oscillations in a room-temperature quantum dot semiconductor optical amplifier (SOA) by shaping the light pulses that trigger them. The experiments described here show that when the excitation is resonant with the short wavelength slope of the SOA gain spectrum, a linear frequency chirp affects its ability to trigger Rabi-oscillations within the SOA: A negative chirp inhibits Rabi-oscillations whereas a positive chirp can enhance them, relative to the interaction of a transform limited pulse. The experiments are confirmed by a numerical calculation that models the propagation of the experimentally shaped pulses through the SOA. © 2016 Optical Society of America**

## 1. Introduction

Triggering and controlling quantum coherent light-matter interactions are key elements in the fundamental study of quantum mechanical systems and their potential applications for quantum information processing, communication, control of chemical reactions, sensing and quantum simulations. Most often, they are observed in studies of isolated systems operating at cryogenic temperatures, such as (cold) atoms or molecules in gaseous phase [1,2], excitons in semiconductor quantum dots (QDs)[3], electron spins in different solids [4,5], and light harvesting molecular complexes [6,7]. Semiconductor opto-electronic devices offer major practical benefits, like room-temperature operation, electrical drive, spectral tunability, and tenability. However, practical opto-electronic applications of such devices usually do not consider any coherent light-matter interactions. The reason is the sub-picosecond coherence time of semiconductors at room temperature [8,9], which impedes any application trying to combine coherent effects with slower, classical, dynamics.

Seeking to bridge the gap between fundamental and applicative studies, we rely on several recent demonstrations of quantum coherent Rabi-oscillations in room-temperature nano-structured semiconductor optical amplifiers (SOAs) operating both at 1.55 µm [10, 11] and 1.3 µm [12]. These observations made use of short pulse excitations and ultrashort characterization techniques and offer opportunities to study the short lived quantum-mechanical interface of radiation and a QD medium in those conditions. Hence, they impact future classical as well as quantum mechanical applications.

The work we report here goes beyond demonstrations of Rabi-oscillations in QD systems, and aims to control them, by shaping the excitations used to trigger them. The shaping is performed by modulating the spectral phase (SP) profiles of the excitation pulses. This re-orders the appearance of the different spectral components of the pulses and allows driving the interaction along a pulse in a desired manner. Following the shaping, pulses are launched resonantly into an InAs/InP QD SOA. The time-dependent complex field envelopes emerging at the SOA output port are characterized by cross frequency resolved optical gating (XFROG) [13] and compared to their input shapes. The observed modifications are interpreted to identify the light-matter interactions experienced by the pulses while propagating along the SOA. This interpretation is supported by a comprehensive numerical model of the pulse propagation within the SOA, accounting for the electronic dynamics in the QD system [11,14], as well as for nonlinear pulse propagation effects [15]. This technique enables to explore details of the coherent light-matter interactions in the presence of incoherent non-resonant propagation mechanisms in the SOA waveguide.

Coherent control experiments in different systems have been utilizing a variety of pulse shapes, ranging from simple linear frequency chirps [3], through triangular or sinusoidal phase modulation [16,17], to feedback generated pulse shapes [18]. Due to the complexity of the SOA dynamics, we chose to demonstrate here the simplest experiment examining the effect of a quadratic SP (QSP) profile that spreads the pulse spectral components linearly in time. It confirms a recent numerical prediction [19] showing that this linear chirp has a major effect on the Rabi-oscillations triggered by such pulses; when the pulse is spectrally located on the short wavelength slope of the SOA gain curve, a positive linear chirp results in pronounced Rabi-oscillations, while a negative linear chirp diminishes them.

## 2. Experimental system

The experimental setup is schematically sketched in Fig. 1 (a). The pulses are generated by a Toptica FemtoFiberPro IR laser, filtered to obtain pulses of about 160 fs duration (when the pulses are shaped to transform limited (TL) envelopes) with a bandwidth of 35 nm centered around 1550 nm. They are split into two arms: a reference pulse, that serves the XFROG measurement, and an excitation pulse, that is fed to the pulse-shaping system. The shaped pulses are launched with controlled intensities at the input facet of the SOA, which is operated under controlled bias and temperature conditions. The pulses at the output of the SOA are coupled to the XFROG system for characterization.

The pulse-shaping system is shown in the front of Fig. 1 (a). It is based on a spatial light modulator (SLM) which is placed in the Fourier plane of a 4f optical imaging system [20]. The SLM (Jenoptiks SLM-S640 with broadband antireflective coating) has a single array of liquid crystal pixels and therefore allows manipulating either the SP profile or the spectral intensity, but not both. To avoid power losses, we chose to alter only the SP profile of the pulses. Two holographic gratings with a resolution of 900 lines /mm, and two curved mirrors with focal length of 200 mm make up the 4f system, and provide the system with a spectral resolution of about 25 GHz/pixel. Operation at 1550 nm poses two challenges; firstly, the phase delay swing of the SLM in this band is only slightly larger than 2π, attributing to some non-idealities of the shaping process, such as phase-folding induced pulse replica [21]. Second, the dispersion power of available gratings in this wavelength range is low yielding sub-optimizing spectral resolution of the SLM.

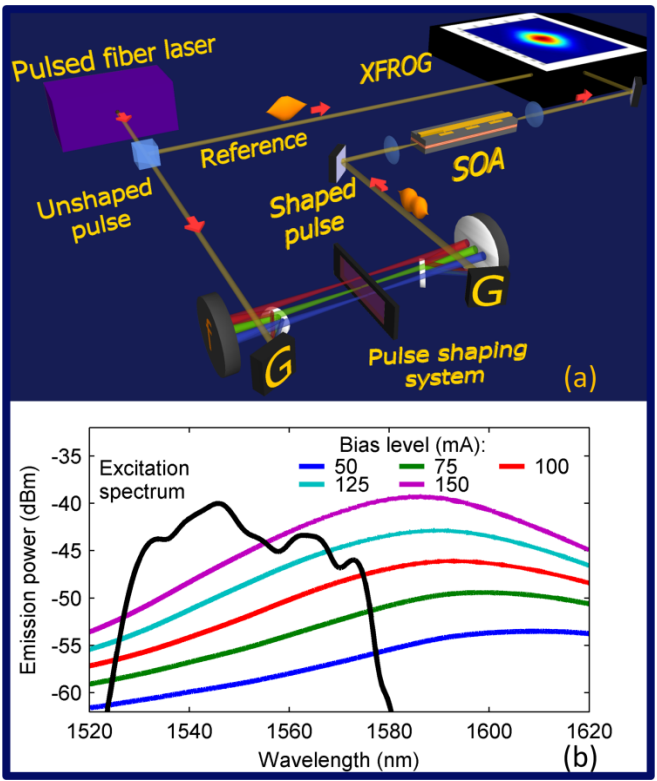

Fig. 1. (a) Schematic of the experimental setup including the pulse-shaping system. (b) Bias dependent electroluminescence spectra of the SOA, together with the spectrum of the excitation pulses used in the experiments.

In the SOA, light is guided along 1.5 mm single mode ridge waveguide. The active region of the SOA comprises six InAs/InAlGaAs QD layers embedded between InAlAs cladding layers and grown on InP substrate (more details are found in Ref. [22]). Its bias dependent electroluminescence spectra are shown in Fig. 1 (b). At high bias levels, these spectra represent approximately the gain spectrum of the SOA. The black plot in Fig. 1 (b) shows the excitation pulses intensity spectrum (which is indifferent to the SP mask encoded by pulse-shaping system). As shown, the pulses are resonant with the short wavelength slope of the gain spectrum. General modelling of the experiments at hand, discussed in [19], predicts that this spectral configuration plays a major role in the control of the coherent Rabi oscillations.

The details of our XFROG system are discussed in details in previous publications. [11]. In the present experiments we use it to characterize both the pulse shapes emerging at the output of the SOA and the designed input pulse shapes (using a bypassing beam path not shown in Fig. 1 (a)). The measurements enable the identification of different processes affecting the output pulse, including Rabi-oscillations, and hence to observe their evolution as the input pulse shape is modified. The results are compared to additional calculations done with the numerical model, where the actual, measured, input pulse shapes are used as excitations rather than the ideal Gaussian pulses used in Ref. [19].

## 3. Experimental results

The simplest shape that can be induced using the present pulse-shaping system is a monotonic pseudo-linear frequency chirp along the pulses. This is obtained by encoding a QSP profile whose curvature is the tuning parameter. Measured pulse envelopes are presented for different QSP curvatures in the left column of Fig. 2, showing their time dependent intensity and instantaneous frequency profiles. They are stretched in time by the induced QSP as different spectral components of the pulses separate from each other. Unlike the well-known case of a Gaussian pulse, the real pulses used here do not acquire a perfect linear chirp since their initial spectrum is rather irregular (see Fig. 1 (b)). Furthermore, the parasitic non-idealities of the shaping system in the 1550 nm wavelength range contribute additional deviations from an ideal linear chirp profile. Nevertheless, the desirable chirp trends are visible; Negative and positive QSP curvatures lead to down or up going chirp slope, respectively.

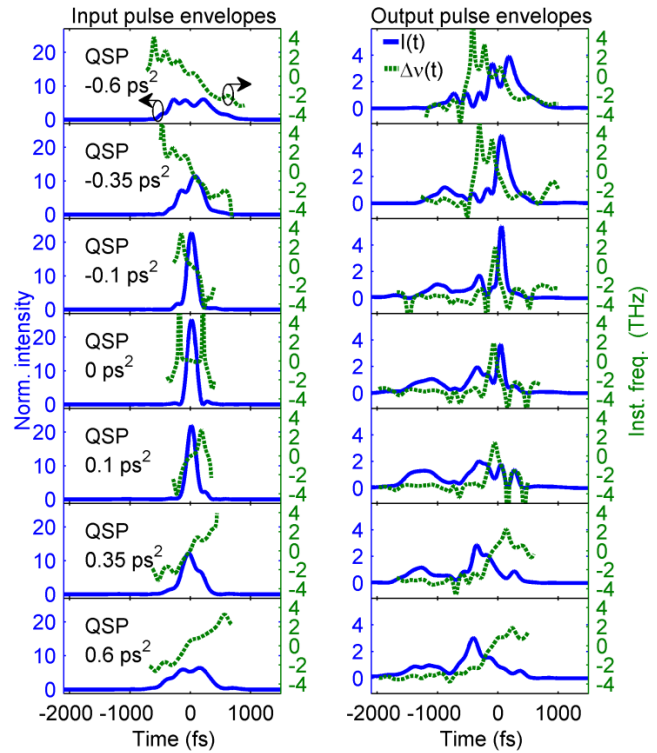

Fig. 2. Input (left) and output (right) pulse envelopes for various QSP values. Continuous blue plots show time dependent intensity profiles, normalized relative to a uniform value. Dashed green curves show the profiles of instantaneous frequency deviations from the central pulse frequency.

The corresponding output pulse envelopes for a 150 mA SOA bias (high gain regime) and input pulse energy of 180 pJ are presented in the right column of Fig. 2, demonstrating the effect of the QSP on their propagation through the SOA. All pulses were measured with respect to the same time axis, enabling a comparison of their features accordingly. As the pulses are spectrally located on the short wavelength slope of the SOA gain spectrum, their lower frequency components are amplified more significantly. Therefore, negatively chirped input pulses interact strongly with the gain medium at a late time during the pulse, as demonstrated by their delayed peaks at the output. The strong interaction occurs earlier as the QSP parameter, and the chirp slope, increase towards positive values. Accordingly, the peak of the output pulses advances to earlier times as the QSP curvature is tuned upwards.

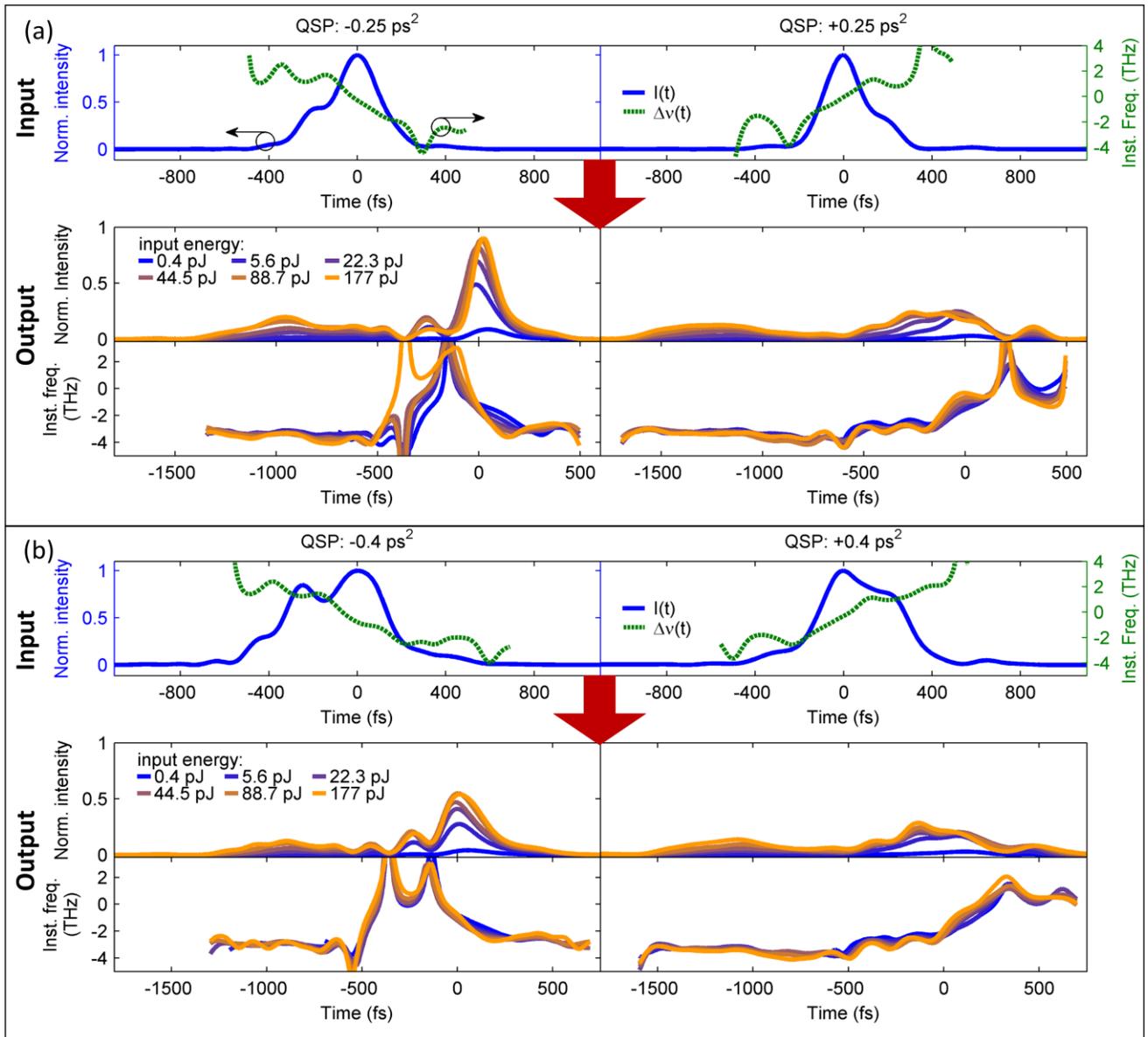

Fig. 3. Output vs. input pulse envelopes with different QSP levels and for varying input pulse energies. (a) QSP level of -0.25 ps$^2$ (left column) and +0.25 ps$^2$ (right column). (b) QSP level of -0.4 ps$^2$ (left column) and +0.4 ps$^2$ (right column). In both sub-figures, the upper row panels show the input envelopes, with time dependent intensity in blue, and instantaneous frequency in dashed green. The middle panels show the output intensity profiles, and the bottom panels show the corresponding instantaneous frequency profiles.

Input pulse energy dependent responses for QSP curvatures of -0.25, and +0.25 ps$^2$ are shown in Fig. 3 (a). In this display, only pulses having the same initial shape were placed on the same time axis. The comparison between the different pulse shape cases is based on the evolution of the features imprinted on them as their input intensity rises. Common to all the pulses is a low frequency and low amplitude leading lobe appearing near t=-1000 fs. It originates from some unshaped features of the input pulses which are amplified. Around the main lobes, the different pulses exhibit significant differences that depend on the QSP value. The negatively chirped pulses show a small leading lobe at t=-300 fs, and a sharp main lobe at t=0 fs. As their input energy is raised, the main lobe compresses and the chirp slope steepens. In contrast, the positively chirped pulses emerge from the SOA broadened, without any separate leading lobe. The peak intensity appears at earlier times as their energy increases, while their trailing edge (around t=0 fs) exhibits absorption accompanied by a rising peak in the instantaneous frequency. Based on the experience gained in previous experiments with similar SOAs [10,11], this feature is identified unequivocally as a Rabi-flop. Namely, these results are consistent with our previous numerical prediction [19].

Results of a similar series of experiments, with QSP curvatures of -0.4, and +0.4 ps$^2$ and the same input energies, are shown in Fig. 3 (b). As the absolute value of the QSP curvature is raised, the pulses broaden with respect to the pulses in Fig. 3 (a). This makes the signatures they exhibit milder, consistent with our predictions [19] stating that a sufficiently large chirp ultimately hinders the appearance of Rabi-oscillations.

The transmitted pulse shapes for TL input pulses (with the same input energies) are presented in Fig. 4. They reveal a mixture of the two types of behaviors described above. Although the main lobe carries no oscillations on its peak, a leading lobe splits away and a notch is building up as the intensity increases, advancing energy from the main lobe towards the leading part of the pulse. As this process is accompanied by a sharp rise of the instantaneous frequency profile, it also resembles the signature of a Rabi-oscillation. On the other hand, Rabi-oscillations are not expected at the leading edge of the pulse without being followed by additional oscillations near or at the intensity peak. Moreover, this pattern is inconsistent with the numerical predictions that predict less pronounced marks of Rabi-oscillations on TL pulses, as compared with slightly positively chirped pulses. This discrepancy is addressed in the discussion section, below.

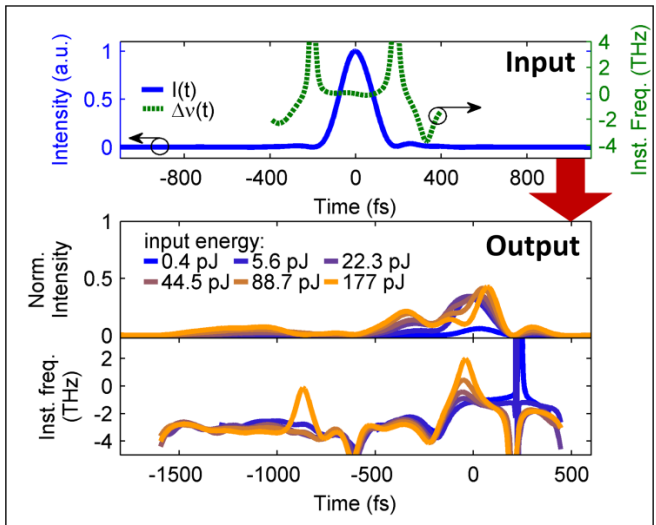

Fig. 4. Output vs. input pulse envelopes for a TL input pulse, for various input pulse energies. The upper panel shows the input envelope, with time dependent intensity in blue, and instantaneous frequency in dashed green. The middle panel shows the output intensity profiles, and the bottom panel shows the corresponding instantaneous frequency profiles.

## 4. Discussion

In order to clarify the differences between the numerical predictions (which use ideal Gaussian pulses [19]) and the experiments, the calculations were repeated except that the model was fed with the measured input pulse shapes, of QSP values -0.25, 0, and 0.25 ps$^2$. This resulted in a highly accurate reproduction of the experimentally observed patterns at the output of the SOA, and allowed to reveal their origins.

The pulses were centered at 1550 nm wavelength, whereas the gain spectrum peaked at 1580 nm, and was about 40 nm wide (FWHM).

The model assumes a short (0.3 mm) propagation length in the SOA, to reduce the computational effort. The interaction parameters and pulse intensities at the input are therefore modified artificially to the level that ensures a match of the calculated output pulse shapes with the shape of the measured ones and their evolutions. Most of the parameter values used here are presented in our previous publications concerning this model [15,19]. Here, the parameters that yield the best reproductions are: a group-velocity dispersion of 30000 fs$^2$/mm, a two-photon absorption (TPA) coefficient of 30 cm/GW and an alpha parameter [23] of -0.6 representing the Kerr effect that accompanies TPA [15,23]. The calculated output pulse shapes are presented in Fig. 5 (a), (b), and (c), for the negatively chirped, TL, and positively chirped input pulses, respectively, each with input energies ranging from 10 pJ to 650 pJ. The calculation corresponds very well with the measured pulse shapes and their trends with increasing intensities. It somewhat overestimates the effect of the Rabi-oscillations on the positively chirped pulse, but reproduces well the signatures on top of both the TL case and the negatively chirped pulses. Thus, it enables to peer deeper into the interaction process.

The evolution of the pulse within the SOA can be studied also by analyzing the occupation probabilities of the two-level systems during their interaction with the pulse. Such an analysis is described in Fig. 6, for the interaction of differently shaped pulses having an input energy of 650 pJ. Shown in Fig. 6 are snapshots of the population inversion probability spectra (the difference between the upper level and lower level occupation probabilities across the spectrum) at different instances of the pulses' propagation along the SOA. The top row in Fig. 6 describes the evolution of the interaction with the negatively chirped pulse, the middle row represents the TL case, and the bottom row describes the positively chirped case. In those snapshots, red shades indicate positive population inversion (gain) due to the applied bias, while dark blue shades indicate absorption caused by floppings of the population that result from the coherent interaction with the pulses. A careful analysis of the leading edge of these interaction images enables to decipher the origin of the various degrees of pulse splitting seen on the leading edges of the differently chirped pulses at the output of the SOA.

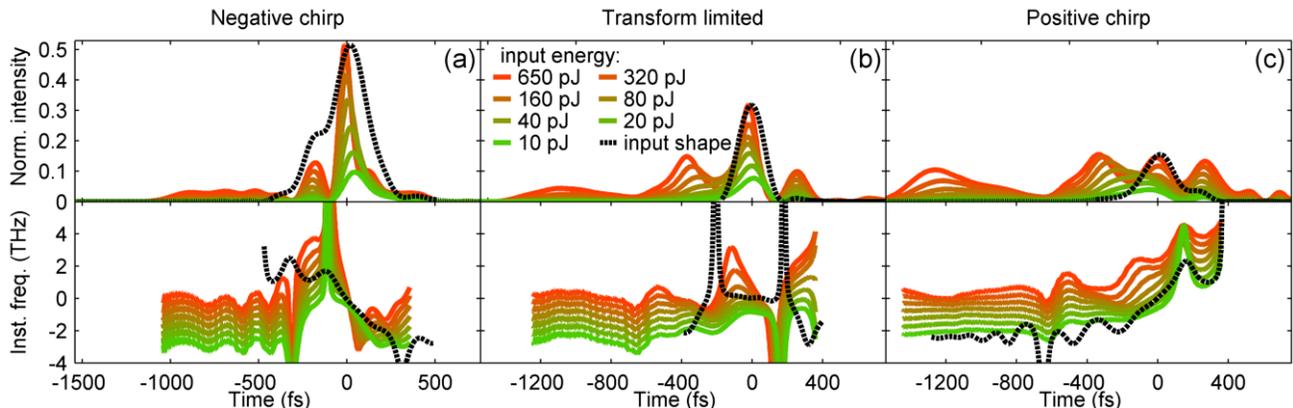

Fig. 5. Calculated output pulse shapes for the experimentally measured input pulse shapes with various input energies. Top row panels present time dependent intensity profiles, while bottom row panels show the corresponding instantaneous frequency profiles. (a) QSP=-0.25 ps$^2$ (b) TL case: QSP=0 (c) QSP=+0.25 ps$^2$.

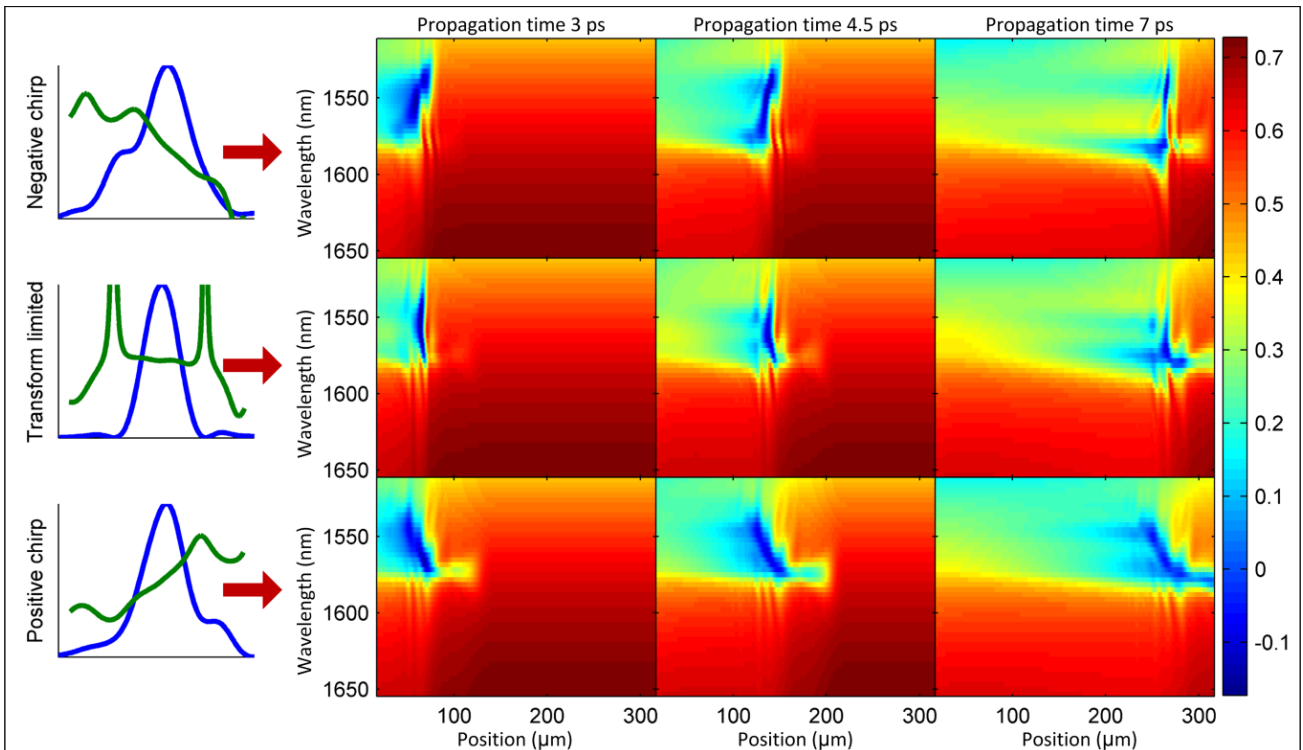

Fig. 6. Calculated snapshots of population inversion probability spectra in the SOA after 3, 4.5, and 7 ps of pulse propagation, for the three types of pulse shapes (shown in the left most column), with 650 pJ input energy. Top row: Negatively chirped pulse with QSP=-0.25 ps$^2$. Middle row: Transform limited pulse with QSP=0. Bottom row: Positively chirped pulse with QSP=+0.25 ps$^2$.

In all cases, a pre-curser is visible in the long wavelength edge of the interaction spectrum (ca. 1580 nm) that produces a small decrease in the population inversion. This originates from a weak leading signal that is only mildly affected by the shaping process. Nevertheless, the differences in its footprint for the different excitations arise from its different intensities, which are somewhat altered by the shaping procedure. This part of the interaction evolves to create the leading lobe preceding the peak of the pulse by about 1 ps (See Fig. 5).

Following this feature, the population inversion returns rapidly to the inverted state. For the positively chirped pulse, this transition is minor, but as the induced chirp becomes negative, a sharp transition appears, that almost crosses the entire interaction spectrum. This feature testifies to a coherent interference between the imprint of the pre-cursor and that of the leading edge of the main portion of the pulse that precedes it. When these two parts are in phase, such as in the positively chirped pulse, this interference is very mild, but when negative chirp is induced on the pulse, this interference is significant. A similar pattern is visible also immediately after the leading part of the pulse, with the same dependence on the nature of the induced chirp. These interferences in the population inversion also affect the refractive index spatial profile experienced by the pulses, hence modifying their instantaneous frequency profiles.

These features evolve along propagation, and due to the effect of non-resonant light-matter interactions (mainly the Kerr effect accompanying TPA [23] and group-velocity dispersion) are sharpened. They ultimately result in the appearance of small leading lobes that are split by notches from the main lobe of the pulse. For the positively chirped pulses, the interference patterns almost do not exist, and therefore the pulses at the output do not acquire such leading satellite lobes. For the negatively chirped pulses, the interference pattern is so prominent, that it results in satellite lobes separated by nodes of the field from the main lobe. In the intermediate, TL, case the interference is less pronounced, and hence its appearance depends on its sharpening by the combination with the non-resonant effects. Thus, it explains the formation of the leading lobe, and the transfer of energy towards it following an increase in the input pulse energy. As such, this signature does not represent Rabi-oscillations.

To complete the confirmation of our earlier predictions, we verify that indeed Rabi-oscillations induce the signatures imprinted on the positively chirped pulses. To that end, observing the bottom row in Fig. 6 reveals that the population inversion along the main part of the interaction with that pulse evolves in a non-monotonous manner. Right from the beginning of the propagation it shows spectral side tails that indicate the Rabi-oscillations. Such oscillations are seen, but with much less clarity, as the chirp induced on the pulses tends to the negative side.

## 5. Conclusion

To conclude, we have demonstrated the ability to modify the coherent interaction between an ultra-short, shaped, pulse and the interband transition in a room-temperature QD SOA. The control procedure employs frequency chirping of the pulses determined by the level of induced QSP. The results we presented are for pulses that are resonant with the short wavelength slope of the SOA gain spectrum; the pulse spectral placement plays an important role. We find that a negative chirp slope inhibits Rabi-oscillations, whereas a positive chirp slope enhances their signature, as compared with the case of a TL pulse, confirming an earlier theoretical prediction. At the same time, it was shown how different coherent features in the excitation pulse shape (such as interference patterns) can affect the interaction of the pulse with the SOA, due to a combination of partly coherent and non-resonant light-matter interactions. These points were illustrated in the measurements, utilizing a pulse-shaping apparatus designed for minimal power loss while operating at 1.55 μm wavelength. The measurements were confirmed by numerical calculations that consider all the interactions experienced by the propagating pulse and which are fed by the actual measured input pulse envelopes.

The ability to control the coherent interactions opens the way to more sophisticated experiments which will investigate, in a controlled fashion, the nature of the coherence in this system, the means to use it for further applications, and the way in which these ultrafast dynamics can affect the classical applications of QD SOAs.

**Funding**. Israel Science Foundation under grant number 293/11.

**Acknowledgment**. O. K. wishes to thank Prof. Atac Imamoglu for providing a new perspective for this work. O. K. thanks also the financial support from the Adams Fellowship, of the Israeli Academy of Sciences and Humanities. A. K. M acknowledges financial support of the PBC Program for Fellowships for Outstanding Post-doctoral Fellows from China and India.